\input{aipcheck}

\documentclass[
  ,draft            
  ]
  {aipproc}

\layoutstyle{6x9}

\usepackage{amsfonts} 
\usepackage{amssymb}
\usepackage{bbm}
\usepackage{epsfig}
\usepackage{graphics}
\usepackage{graphicx}
\usepackage{subfigure}

\def\ptps{{Prog.\ Theor.\ Phys.\ Suppl. \ }}

\newcommand{\gsim}{\,\lower2truept\hbox{${>\atop\hbox{\raise4truept\hbox{$\sim$}}}$}\,}
\newcommand{\pp}{~~~.}
\newcommand{\vv}{~~~,}

\newcommand{\be}{\begin{equation}}
\newcommand{\ee}{\end{equation}}
\newcommand{\bea}{\begin{eqnarray}}
\newcommand{\eea}{\end{eqnarray}}

\begin{document}

\title{Clustering in growing neutrino cosmologies}


\author{Valeria Pettorino}{
  address={Institut  f\"ur Theoretische Physik, Universit\"at Heidelberg,
Philosophenweg 16, D-69120 Heidelberg}
}

\author{David F. Mota}{
  address={Institut  f\"ur Theoretische Physik, Universit\"at Heidelberg,
Philosophenweg 16, D-69120 Heidelberg}
}

\author{Georg Robbers}{
  address={Institut  f\"ur Theoretische Physik, Universit\"at Heidelberg,
Philosophenweg 16, D-69120 Heidelberg}
}

\author{Christof Wetterich}{
  address={Institut  f\"ur Theoretische Physik, Universit\"at Heidelberg,
Philosophenweg 16, D-69120 Heidelberg}
}

\begin{abstract}
We show that growing neutrino models, in which the growing neutrino mass stops the dynamical evolution of a dark energy
scalar field, lead to a substantial neutrino clustering on the scales of
superclusters. Nonlinear neutrino lumps form at redshift $z \approx 1$ and
could partially drag the clustering of dark matter. If observed, 
large scale non-linear structures could be an indication for a new attractive 
force stronger than gravity and mediated by the `cosmon' dark energy scalar field.
\end{abstract}

\maketitle


Realistic
quintessence models need to explain the recent increase of $\Omega_{\phi}(z)$ by a
factor of around ten or more \cite{wetterich_1988, ratra_peebles_1988, barreiro_etal_2000, ferreira_joyce_1998}. It has been recently proposed
\cite{amendola_etal_2007, wetterich_2007, mota_etal_2008} that a growing mass of the neutrinos
may play a key role in stopping the dyna\-mical evolution of the dark energy
scalar field, the cosmon. For a slow evolution of the cosmon, the scalar
potential acts like a cosmological constant, such that the equation of state
of dark energy is close to $w = -1$ and the expansion of the universe
accelerates. In these models, the onset of accelerated expansion is triggered
by neutrinos becoming non relativistic. For late cosmology, $z \gsim 5$, the
overall cosmology is very similar to the usual $\Lambda$CDM concordance model with a cosmological constant.

An efficient stopping of the cosmon evolution by the relatively small energy
density of neutrinos needs a cosmon-neutrino coupling that is somewhat larger
than gravitational strength. This is similar to mass varying neutrino
models \cite{fardon_etal_2004, afshordi_etal_2005, gu_etal_2003, bjaelde_etal_2007, peccei_2005, takahashi_tanimoto_2006, weiner_zurek_2006, das_weiner_2006, bi_2005, schwetz_2006, spitzer_2006, fardon_etal_2006, li_etal_2006, hung_2000, brookfield_etal_2007, brookfield_etal_2006}, even though the coupling in those models is generically much larger.
In turn, the enhanced attraction between
neutrinos leads to an enhanced growth of neutrino fluctuations, once the
neutrinos have become non-relativistic \cite{afshordi_etal_2005,bjaelde_etal_2007,amendola_etal_2007,bean_etal_2007}. In view of
the small present neutrino mass, $m_\nu(t_0) < 2.3 eV$, and the time
dependence of $m_\nu$, which makes the mass even smaller in the past, the time
when neutrinos become non-relativistic is typically in the recent history of
the universe, say $z_R \approx 5$. Neutrinos have been free streaming for $z >
z_R$, with a correspondingly large free streaming length. Fluctuations on
length scales larger than the free streaming length are still present at
$z_R$, and they start growing for $z < z_R$ with a large growth rate. This
opens the possibility that neutrinos form nonlinear lumps
\cite{amendola_etal_2007, brouzakis_etal_2007} on supercluster scales, thus
opening a window for observable effects of the growing neutrino scenario.

We show that neutrino perturbations indeed grow non-linear in
these models. Non-linear neutrino structures form at redshift $z \approx 1$ on
the scale of superclusters and beyond. One may assume that these structures
later turn into bound neutrino lumps of the type discussed in
\cite{brouzakis_etal_2007}. 
Our investigation is limited, however, to linear perturbations. We can therefore provide a reliable estimate for the time when the first fluctuations become non-linear. For later times, it should only be used to give qualitative limits.

A crucial ingredient in this model is the dependence of the neutrino mass on
the cosmon field $\phi$, as encoded in the dimensionless cosmon-neutrino
coupling $ \beta(\phi) \equiv - \frac{d \ln{m_\nu}}{d \phi}.$ For increasing $\phi$ and $\beta < 0$ the neutrino mass increases with
time $ m_{\nu} = \bar{m}_{\nu} e^{-{\tilde{\beta}}(\phi) \phi},$ where
$\bar{m}_{\nu}$ is a constant and $\beta = \tilde{\beta} + \partial
\tilde{\beta}/\partial \ln{\phi}$. The coupling $\beta(\phi)$ can be either a
constant \cite{amendola_etal_2007} or, in general, a function of $\phi$, as
proposed in \cite{wetterich_2007} within a particle physics model. The cosmon field $\phi$ is normalized in units of the reduced Planck mass $M = (8 \pi G_N)^{-1/2}$, and $\beta \sim 1$ corresponds to a cosmon mediated interaction for neutrinos with gravitational strength. For a given
cosmological model with a given time dependence of $\phi$, one can determine
the time dependence of the neutrino mass $m_\nu(t)$. For three degenerate
neutrinos the present value of the neutrino mass $m_\nu(t_0)$ can be related
to the energy fraction in neutrinos via $\Omega_{\nu} (t_0) = (3 m_\nu
  (t_0))/(94\, eV h^2)$. 

We choose
an exponential potential \cite{wetterich_1988, ratra_peebles_1988, barreiro_etal_2000, ferreira_joyce_1998}:
\be \label{pot_def} V(\phi) = M^2 U(\phi) = M^4 e^{- \alpha \phi} \vv \ee
where the constant $\alpha$ is one of the free parameters of our model.

The homogeneous energy density and pressure of the scalar field $\phi$ are defined
in the usual way \cite{mota_etal_2008}. 
We can express the conservation equations for dark energy and growing
matter in the form of neutrinos as \cite{wetterich_1995, amendola_2000} 
\bea \label{cons_phi} \rho_{\phi}' = -3 {\cal H} (1 + w_\phi) \rho_{\phi} +
\beta(\phi) \phi' (1-3 w_{\nu}) \rho_{\nu} \vv \\
\label{cons_gr} \rho_{\nu}' = -3 {\cal H} (1 + w_{\nu}) \rho_{\nu} - \beta(\phi) \phi' (1-3 w_{\nu}) \rho_{\nu}
\pp \nonumber \eea 
The sum of the energy momentum tensors for neutrinos and the cosmon is conserved,
 but not the separate parts. We neglect a possible cosmon coupling
to Cold Dark Matter (CDM), so that $\label{cons_cdm} \rho_c' = -3 {\cal H} \rho_c $.

\begin{figure}[ht]
\begin{picture}(185,200)(0,0)
\includegraphics[width=75mm,angle=0.]{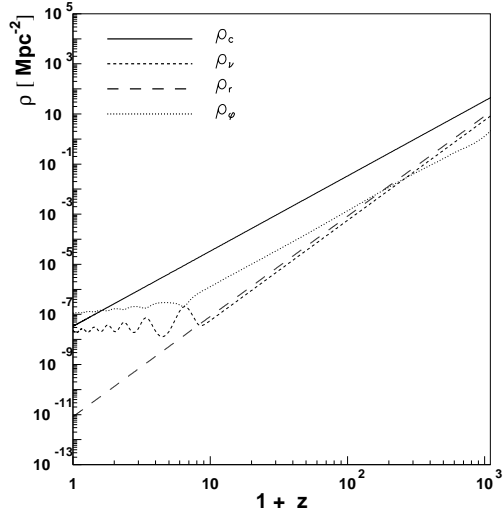}
\end{picture}
\caption{Energy densities of neutrinos (dashed), cold dark matter (solid), dark energy (dotted) and photons (long dashed) are plotted vs redshift. For all plots we take a constant $\beta = -52$, with $\alpha = 10$ and large neutrino mass $m_\nu = 2.11 \, eV$.}
\label{fig_1}
\vspace{0.5cm}
\end{figure}

For a given potential (\ref{pot_def}) the evolution equations for the different species can be numerically integrated, giving the background
evolution shown in FIG.\ref{fig_1} (for constant $\beta$) \cite{amendola_etal_2007}. The initial
pattern is a typical early dark energy model, since neutrinos are still relativistic and almost
massless. Radiation dominates until matter radiation equality, when CDM takes over. Dark
energy is still subdominant and falls into the attractor provided by the
exponential potential (see \cite{wetterich_1995, amendola_2000} for
details). As the mass of the neutrinos increases with time, the term
$\sim \beta \rho_\nu$ in the evolution equation for the cosmon (\ref{cons_phi})
starts to play a more significant role, kicking
$\phi$ out of the attractor as soon as neutrinos become non-relativistic. This resembles the effect of the coupled dark matter component in \cite{huey_wandelt_2006}. Subsequently, small decaying oscillations characterize the $\phi - \nu$ coupled fluid
and the two components reach almost constant values. The va\-lues of the energy densities
today are in agreement with observations, once the precise crossing time for the end of the scaling 
solution has been fixed by an appropriate choice of the coupling $\beta$. At
present the neutrinos are still subdominant with respect to CDM, though in the future
they will take the lead (see
\cite{amendola_etal_2007} for details on the future attractor solution for constant $\beta$).

\begin{figure}[ht] 
\begin{minipage}{70mm}
\begin{center}
\begin{picture}(185,200)(20,0)
\includegraphics[width=78mm,angle=0.]{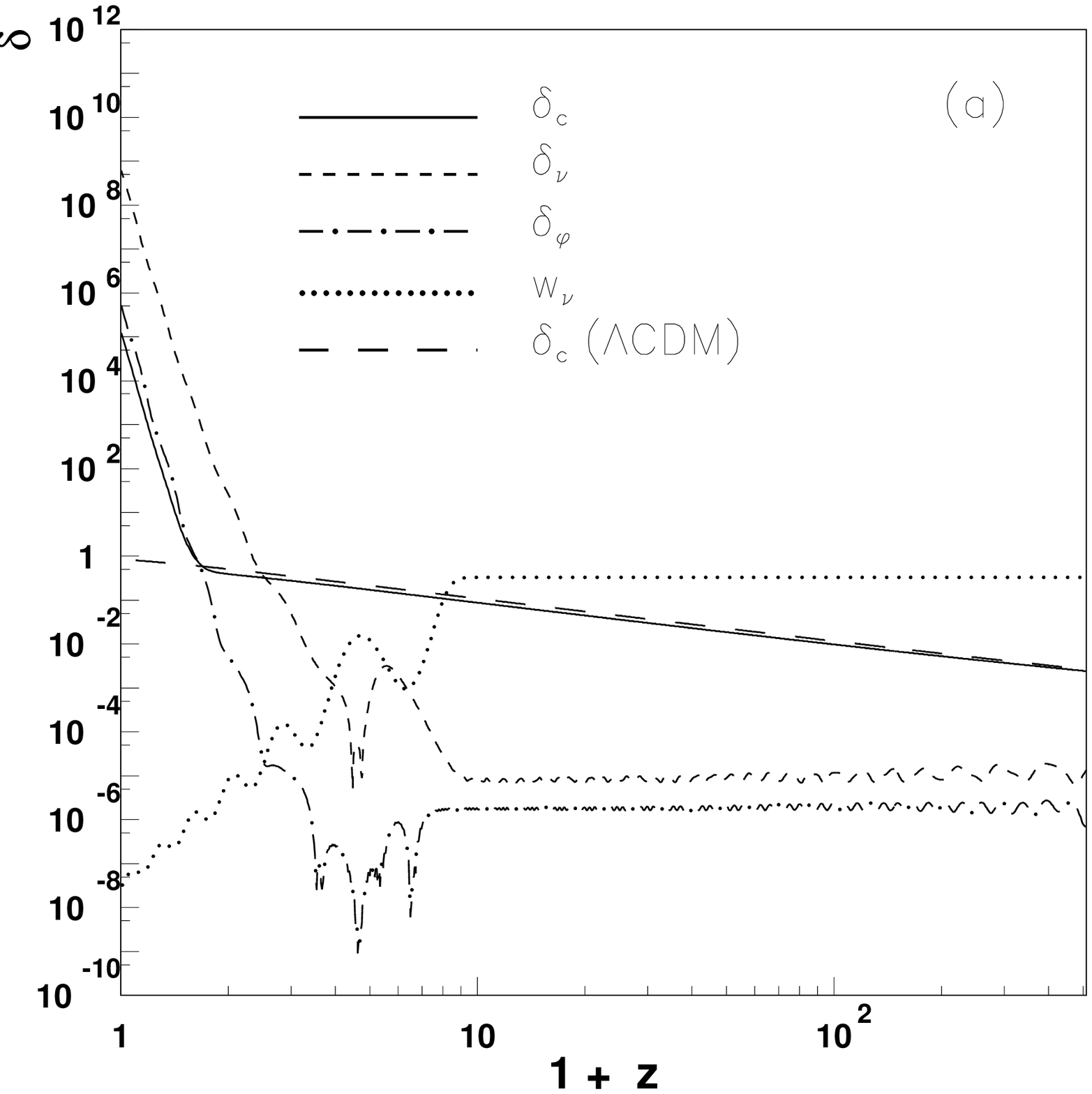}
\end{picture}
\label{fig_2a}
\end{center}
\end{minipage}
\hspace{4mm}
\begin{minipage}{70mm}
\begin{center}
\begin{picture}(185,200)(20,0)
\includegraphics[width=78mm,angle=0.]{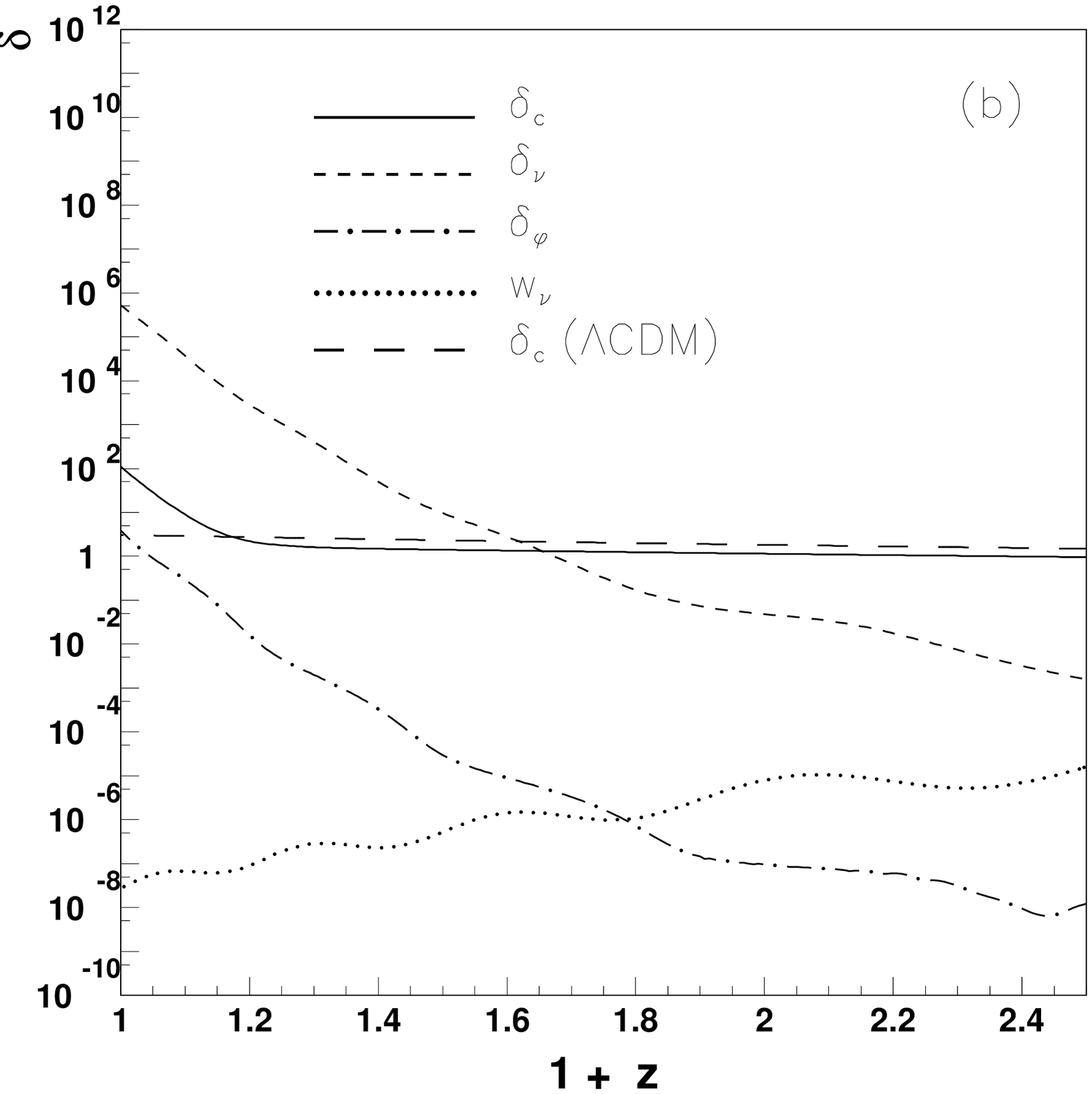}
\end{picture}
\label{fig_2b}
\end{center}
\end{minipage}
\vspace{4mm}
\caption{Longitudinal density perturbation for CDM (solid), $\nu$ (dashed) and $\phi$ (dot-dashed) vs redshift for $k = 0.1 h/Mpc$ (upper panel) and $k = 1.1 h/Mpc$ (lower panel, $\lambda = 8 Mpc$). The neutrino equation of state (dotted) is also shown. The long dashed line is the reference $\Lambda$CDM.}\label{fig_2}
\end{figure}

The evolution equations for linear perturbations (in Fourier space), in Newtonian gauge (in which the non diagonal metric perturbations are fixed to zero) \cite{kodama_sasaki_1984}, is fully described in \cite{mota_etal_2008}.

We numerically compute the linear density
perturbations both using a modified version of CMBEASY \cite{cmbeasy} and, independently, a modified version of CAMB \cite{lewis_etal_2000}. We plot the density fluctuations $\delta_i$ as a function of
redshift for a fixed $k$ in FIG.\ref{fig_2}.
The neutrino equation of state is also shown, starting from
$1/3$ when neutrinos are relativistic and then decreasing to its present value
when neutrinos become non relativistic. The turning point marks the time at
which neutrino perturbations start to increase. At the scale of $k = 0.1 h/Mpc$
(corresponding to superclusters scales) shown in FIG.\ref{fig_2}$a$, neutrino
perturbations eventually overtake CDM perturbations and even force $\phi$
perturbations to increase as well, in analogy with dark 
energy clustering expected in \cite{perrotta_baccigalupi_2002, pettorino_baccigalupi_2008} within scalar tensor 
theories. Notice, however, that the scale at which
neutrinos form nonlinear clumps depends on the model parameters, in particular
the coupling $\beta$, the potential parameter $\alpha$ and
the present days neutrino mass. Those are related to the neutrino
free-streaming length, the range of the cosmon
field and its mass. A detailed investigation of the parameter space will be performed in future work, but we mention that the model \cite{wetterich_2007} with varying $\beta$ gives qualitatively similar results.

\begin{figure}[ht]
\includegraphics[width=75mm,angle=0.]{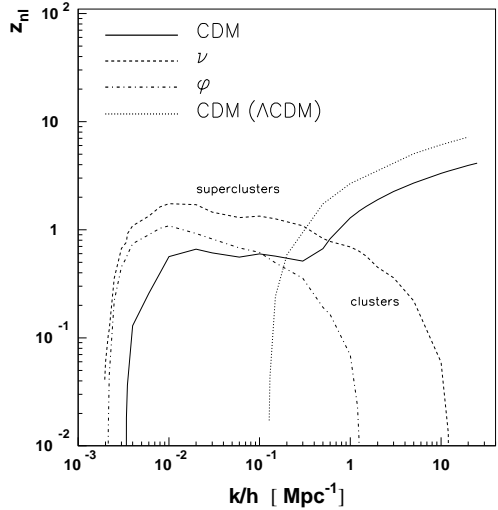}
\caption{Redshift of first non linearities vs the wavenumber $k$ for CDM (solid), $\nu$ (dashed) and $\phi$ (dot-dashed). We also plot CDM for a reference $\Lambda$CDM model (dotted).\label{fig_4}}
\end{figure}

We emphasize again that once the neutrinos form strong nonlinearities - neutrino lumps - one expects that nonlinear effects substantially slow down the increase of $\delta_\nu$ and even stop it. The magnitude of the CDM-dragging by neutrinos is therefore not shown - it might be much smaller than visible in FIG.\ref{fig_2}. These remarks concern the quantitative interpretation of all the following figures, which are always computed in the linear approximation.

Nevertheless, the linear approximation demonstrates well the mechanisms at work. 

In FIG.\ref{fig_4} we plot the redshift $z_{nl}$ at which CDM,
neutrinos and $\phi$  become nonlinear as a function of the
wavenumber $k$. The case of CDM in the concordance $\Lambda$CDM model with the same present value of $\Omega_{\phi}$ and for massless neutrinos is also
shown for reference (dotted line). The
redshift $z_{nl}$ roughly measures when nonlinearities first appear by evaluating the
time at which $\delta(z_{nl}) = 1$ 
for each species. The curves in FIG.\ref{fig_4} are obtained in the linear approximation, such that only the highest curves are quantitatively reliable.  This concerns CDM for large $k$ and neutrinos for small $k$. The subleading components are influenced by dragging effects and may be, in reality, substantially lower. 

We can identify four regimes: i) At very big scales (larger than superclusters) the universe is
homogeneous and perturbations are still linear today.
ii) The range of length scales going from 
$14.5$ Mpc to about $4.4 \times 10^3$ Mpc appears to be
highly affected by the neutrino coupling in growing matter scenarios: neutrino
perturbations are the first ones to go nonlinear and neutrinos seem to form
clumps in which then both the scalar field and CDM could fall into. Note that the
effect of the neutrino fluctuations on the gravitational potential induces CDM to
cluster earlier with respect to
the concordance $\Lambda$CDM  model, where CDM is still linear at scales above $\sim$87 Mpc.
iii) For lengths included in the range between
$0.9$ Mpc and $14.5$ Mpc, CDM takes
over. That is in fact expected since neutrinos start to approach the free streaming scale. In this regime CDM drags neutrinos, and this effect may be overestimated in the linear approximation.
Notice that in our model CDM clusters later than it would do in $\Lambda CDM$. 
There are two reasons for this effect. At early times, the presence of a homogeneous component of early dark energy, 
$\Omega_{\phi} \sim 3/\alpha^2$, implies that $\Omega_m$ is somewhat smaller than one and therefore clustering is slower \cite{doran_etal_2001}.
At later times, $\Omega_m$ is smaller than in the $\Lambda CDM$ model since for the same $\Omega_{\phi}$, part of $1-\Omega_{\phi} = \Omega_m + \Omega_\nu$
is now attributed to neutrinos. In consequence massive neutrinos reduce structure at smaller scales when they do not contribute to the clumping. The second effect is reduced for a smaller present day neutrino mass. 
iv) Finally, at very small scales (below clusters), CDM becomes highly non linear and
neutrinos enter the free streaming regime, their perturbations do not growth and remain
inside the linear regime.

Maximum neutrino clustering occurs on supercluster scales and one may ask about observable consequences. First of all, the neutrino clusters could have an imprint on the CMB-fluctuations. Taking the linear approximation at face value, the ISW-effect of the particular model presented here would be huge and strongly ruled out by observations. However, non-linear effects will substantially reduce the neutrino-generated gravitational potential and the ISW-effect. Further reduction is expected for smaller values of $\beta$ (accompanied by smaller $\alpha$). It is well conceivable that realistic models for the growing neutrino scenario lead to an ISW-effect in a range interesting for observations.

A second possibility concerns the detection of nonlinear structures at very large length scales. Such structures can be found via their gravitational potential, independently of the question if neutrinos or CDM source the gravitational field. Very large nonlinear structures are extremely unlikely in the $\Lambda$CDM concordance model. An establishment of a population of such structures, and their possible direct correlation with the CMB-map \cite{rudnick_etal_2007, inoue_silk_2006, samal_etal_2007, jain_ralston_1998, giannantonio_etal_2008}, could therefore give a clear hint for ``cosmological actors'' beyond the $\Lambda$CDM model. For any flat primordial spectrum the gravitational force will be insufficient to produce large scale clumping, which could thus be an indication for a new attractive force stronger than gravity - in our model mediated by the cosmon.

\begin{theacknowledgments}
DFM and VP acknowledge the Alexander von Humbold Foundation. GR acknowledges support by the Deutsche Forschungsgemeinschaft, grant TRR33 ``The Dark Universe".
\end{theacknowledgments}

\bibliographystyle{aipprocl} 

\end{document}